\documentclass[aps,amsfonts,prb,twocolumn,showpacs]{revtex4-1}

\usepackage{subfigure}
\usepackage{textcomp}
\usepackage{graphicx}
\usepackage{amssymb}
\usepackage{amsmath}
\usepackage{bm}
\usepackage{amsmath, amsthm, amssymb}
\usepackage{dsfont}
\usepackage[colorlinks=true,linkcolor=blue,urlcolor=blue,citecolor=blue]{hyperref}
\usepackage{multirow}
\usepackage{url}

\usepackage{pdfpages} 
\makeatletter
\AtBeginDocument{\let\LS@rot\@undefined}
\makeatother

\def\beq{\begin{align}}
\def\eeq{\end{align}}
\def\bea{\begin{align}}
\def\eea{\end{align}}

\begin{document}

\title{Hydrodynamics of operator spreading and quasiparticle diffusion in interacting integrable systems}

\author{Sarang Gopalakrishnan$^1$, David A. Huse$^2$, Vedika Khemani$^3$, and Romain Vasseur$^4$}
\affiliation{$^1$ Department of Physics and Astronomy, CUNY College of Staten Island, Staten Island, NY 10314;  Physics Program and Initiative for the Theoretical Sciences, The Graduate Center, CUNY, New York, NY 10016, USA}
\affiliation{$^2$ Physics Department, Princeton University, Princeton, New Jersey 08544, USA}
\affiliation{$^3$ Department of Physics, Harvard University, Cambridge, MA 02138, USA}
\affiliation{$^4$ Department of Physics, University of Massachusetts, Amherst, MA 01003, USA}

\begin{abstract}
We address the hydrodynamics of operator spreading in interacting integrable lattice models. In these models, operators spread through the ballistic propagation of quasiparticles, with an operator front whose velocity is locally set by the fastest quasiparticle velocity. In \emph{interacting} integrable systems, this velocity depends on the density of the other quasiparticles, so equilibrium density fluctuations cause the front to follow a biased random walk, and therefore to broaden diffusively. Ballistic front propagation and diffusive front broadening are also generically present in \emph{non}-integrable systems in one dimension; thus, although the mechanisms for operator spreading are distinct in the two cases, these coarse grained measures of the operator front do not distinguish between the two cases. We present an expression for the front-broadening rate; we explicitly derive this for a particular integrable model (the ``Floquet-Fredrickson-Andersen'' model), and argue on kinetic grounds that it should apply generally. Our results elucidate the microscopic mechanism for diffusive corrections to ballistic transport in interacting integrable models. 

\end{abstract}

\maketitle

How an initially local perturbation spreads under time evolution is a central question in many-body quantum dynamics. Recently, a general coarse-grained phenomenology for such ``operator spreading'' was proposed for many-body systems with chaotic dynamics; this description was motivated by the ansatz that chaotic systems have essentially random time-evolution, constrained only by locality and a few local conservation laws~\cite{tsunami, casini2016, mezei2017, FawziScrambling, AdamCircuit1, nvh, vrps, kvh, rpv, cdc_long}. In one dimension, the coarse-grained description suggests that operators spread ballistically, with a ``front'' that broadens diffusively~\cite{nvh, vrps}. In chaotic systems, conventional response functions do not diagnose the operator front, since conventional observables relax locally; even in the case of conserved quantities, the autocorrelation function spreads diffusively while the front spreads ballistically~\cite{KimHuse,  KnapScrambling, LuitzScrambling, kvh, rpv}.
Instead, the dynamics of the operator front can be captured by the out-of-time-order commutator (OTOC)~\cite{lo_otoc, ShenkerStanfordButterfly, maldacena2016bound}, which measures the ``footprint'' of the spreading operator:
${C}({ x},t) \equiv \frac{1}{2} \langle [O_0(t), W_{x}]^\dagger [O_0(t), W_{x}] \rangle$, 
where $W_x, O_0$ are local norm-one operators at position $ x$ and $0$, and  the expectation value is taken in a chosen equilibrium ensemble. As $O_0(t)$ spreads in a chaotic system, ${C}(x,t)$ grows to order one inside a ``light cone'' bounded by the propagating front.

Integrable systems have very different dynamics from chaotic ones: they have ballistically propagating quasiparticles and an extensive number of conservation laws~
\cite{Calabrese:2006, PhysRevLett.106.217206, PhysRevLett.110.257203, PhysRevLett.113.117202, PhysRevLett.115.157201, 2016arXiv160300440I, 1742-5468-2016-6-064010,PhysRevB.89.125101, alba2017entanglement}. Thus, one might expect the dynamics of operator spreading in these systems to differ from that in chaotic systems; and, indeed, integrable systems that can be mapped to free fermions have fronts that broaden subdiffusively as $t^{1/3}$~\cite{Platini2005, PhysRevB.97.081111,2018arXiv180305902K,xu2018a,PhysRevB.96.220302}. 
A quasiparticle description also holds for \emph{interacting} integrable systems, so it is tempting to conclude that such systems also have $t^{1/3}$ broadening of the operator front. 

We argue here that \emph{interacting} integrable systems in fact have operator fronts that broaden diffusively, just as in non-integrable systems. In interacting integrable systems, the ballistically propagating quasiparticles also exhibit subleading diffusive spreading~\cite{percus, el2003thermodynamic, el2005, PhysRevB.89.075139, medenjak2017, kormos2017, doyon2017dynamics, PhysRevB.97.081111, spohn2018interacting, denardis2018, sg_ffa, klobas2018a, klobas2018b}. This subleading behavior manifests itself in the shape of an operator near its front~\cite{sg_ffa}. Although operator fronts broaden diffusively in both chaotic and interacting integrable systems, the mechanisms are different: in the latter case, we expect diffusive broadening of conventional response functions as well as OTOCs. 
Our results show that the behavior of the OTOC at and beyond the front does \emph{not} distinguish between non-integrable and interacting integrable systems, despite the qualitatively different mechanisms governing operator spreading in the two cases. Further, while OTOCs for some operators decay to zero behind the front in non-interacting models, signaling a lack of chaos~\cite{motrunich2018}, we argue below that we expect local operators in interacting integrable models to generically have OTOCs that saturate to a nonzero value as the operator ``fills in" behind the front. It is presently unclear whether this saturation value is universal and distinct from the chaotic case.

We quantitatively address operator spreading in interacting integrable systems using a generalized hydrodynamic framework~\cite{Doyon, Fagotti}. To this end, we develop a simple picture of quasiparticle diffusion using kinetic theory, thinking of quantum integrable systems as soliton gases~\cite{solitongases,BBH, sachdev_young,PhysRevLett.78.943}. According to this picture, a quasiparticle experiences random time delays as it propagates, owing to collisions with other quasiparticles, and these random time delays cause diffusion. This picture is illustrated first for a specific integrable model, the Floquet-Fredrickson-Andersen (FFA) model~\cite{bobenko, prosen2016ffa, sg_ffa},
for which we derive explicit closed-form expressions for the diffusion constant.
Our analytic and numerical results for the FFA model are in excellent agreement with one another. We generalize our results to other integrable models; in the general case, our results can be regarded as a simple kinetic-theory perspective on some of the results obtained in Ref.~\cite{denardis2018} using apparently different methods.

\emph{Picture from Kinetic Theory}.---Our main result is a simple quantitative framework for computing quasiparticle diffusion within generalized hydrodynamics (GHD)~\cite{Doyon, Fagotti, SciPostPhys.2.2.014, GHDII, BBH0, BBH,PhysRevLett.119.020602, doyon2017dynamics, solitongases,PhysRevLett.119.195301}. 
One can understand the origin of diffusion as follows. In an interacting integrable system, the velocity of a quasiparticle depends on the densities of other quasiparticles near it. Generically, this relationship is linear: $\partial v_k / \partial \rho_q \neq 0$, where $k$ and $q$ denote quantum numbers (quasimomenta, species, etc.) of quasiparticles. Further, the densities of each type of quasiparticle exhibit $1/\sqrt{\ell}$ fluctuations in a region of length $\ell$. Thus, $v_k$ should vary by an amount $\sim 1/\sqrt{\ell}$ while passing through such a region, and therefore the time a quasiparticle takes to traverse the region will also fluctuate by $\sqrt{\ell}$. This immediately implies diffusive broadening of the quasiparticle front. 

\begin{figure}[tb]
\begin{center}
\includegraphics[width = 0.45\textwidth]{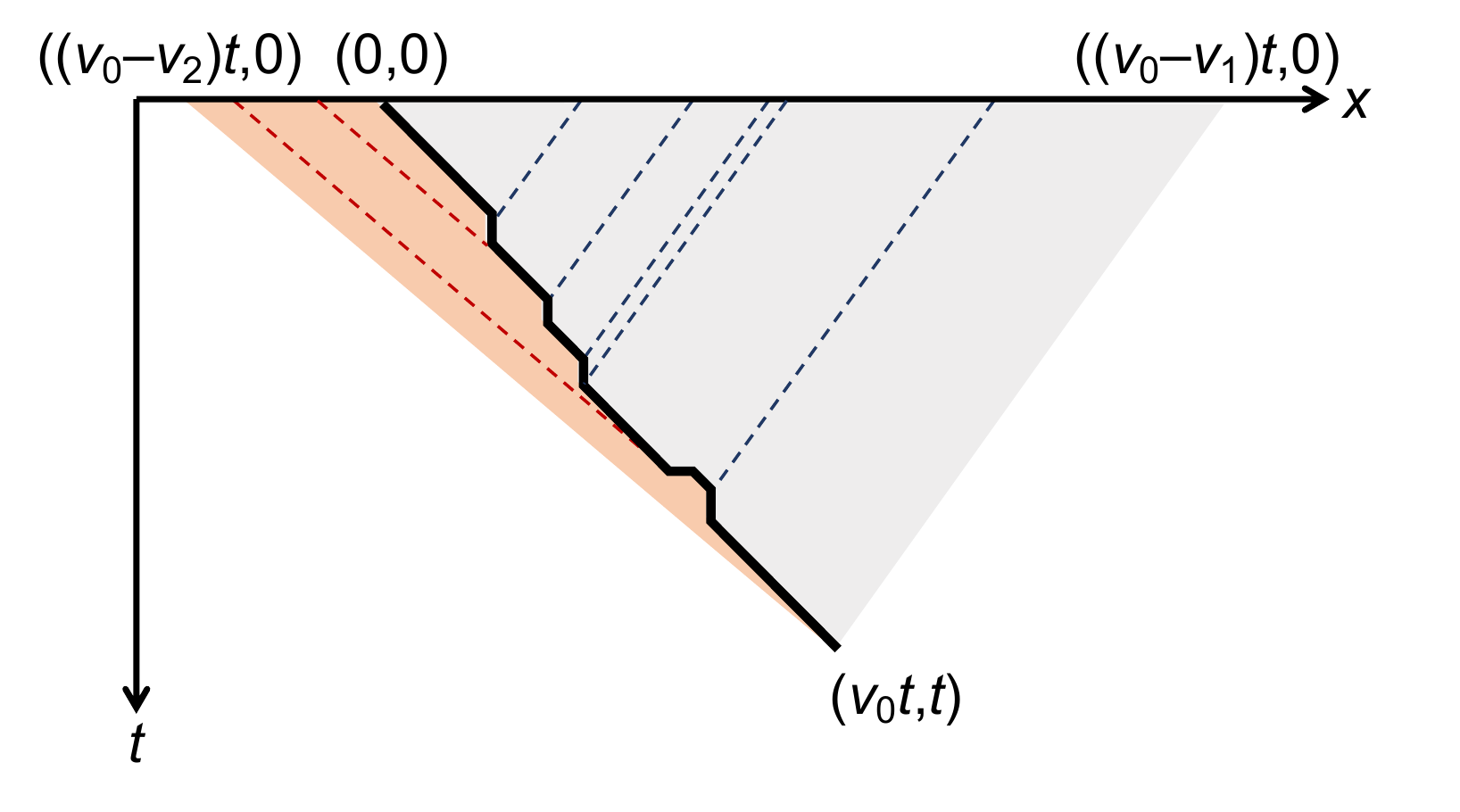}
\caption{Geometric picture of quasiparticle diffusion: the worldline of the ``tagged'' quasiparticle (thick black line), with mean velocity $v_0$, wanders owing to collisions with other quasiparticles (dashed lines). In a time $t$, the tagged quasiparticle collides with quasiparticles moving at velocity $v_i$ so long as those quasiparticles started out in a spatial window of size $|v_i - v_0| t$; their density fluctuations inside this window govern front broadening.}
\label{fig1}
\end{center}
\end{figure}

To develop a more quantitative understanding, we adopt a coarse-grained description of an interacting integrable system in terms of a soliton gas~\cite{solitongases}. To leading order, quasiparticles move ballistically with a speed renormalized by the densities of the other quasiparticles, but they also diffuse because of random shifts due to collisions with other quasiparticles. These shifts are random because of thermal fluctuations. 
Consider a quasiparticle (of type $\alpha$ and group velocity $v_0$) that starts at a position $x = 0$ and travels for a time $t$. In this time interval, it collides with quasiparticles with velocities $v_i$ that were initially at positions between $0$ and  $x_i(v_i) =  (v_0 - v_i) t$ (Fig.~\ref{fig1}). When quasiparticles collide, they scatter elastically; because of these scattering events, the velocity $v_\alpha(k)$ -- with $k$ the pseudo-momentum or rapidity -- depends on the densities of all other quasiparticles. This is the basis of GHD~\cite{Doyon, Fagotti}.

To compute the diffusion of a tagged quasiparticle with quantum numbers $(\alpha, k)$, we account for thermal fluctuations of the densities of the other quasiparticles it collides with. Even though the tagged quasiparticle will move nearly ballistically with velocity $v_{\alpha,k}$, it will also wander owing to collisions with other quasiparticles. In a time $t$, the tagged quasiparticle collides with quasiparticles moving at velocity $v_{\beta,k'}$ if they started out in a spatial window of size $|v_{\alpha,k} -v_{\beta,k'}|t$ (Fig.~\ref{fig1}). The density fluctuations of these quasiparticles govern the diffusive broadening of the ballistic trajectory of the tagged quasiparticle as
\begin{align}\label{front}
\delta x^2_{\alpha,k} (t) = [\delta v_{\alpha,k}]^2 t^2 = t^2 \sum_\beta \int dk' \left(\frac{\partial v_{\alpha,k}}{\partial  n_{\beta,k'}}\right)^2 \!\!\! [\delta  n_{\beta,k'}]^2,
\end{align}
where $ n_{\beta,k'}$ denotes the occupation number (called ``generalized Fermi factor'', to be defined more precisely below) of the quasiparticles of type $\beta$ with pseudo-momentum $k'$. In that formula, we used the fact that the equilibrium fluctuations of the generalized Fermi factor are diagonal~\cite{PhysRevB.54.10845} $\langle \delta  n_{\beta,k'}  \delta  n_{\gamma,k''} \rangle = \delta_{\beta, \gamma} \delta(k' - k'') C_\beta(k')/\ell$, where the fluctuations are computed over a region of size $\ell$. (The $\ell$-dependence is as one would expect from central-limit arguments.) Crucially, the fluctuations of $ n_{\beta,k'}$ are computed over a region of size $\ell =|v_{\alpha,k} -v_{\beta,k'}| t$. Thus, the broadening of the tagged trajectory takes the form
\begin{align}\label{front2}
\delta x^2_{\alpha,k} (t)  = t  \sum_\beta \int dk' \left(\frac{\partial v_{\alpha,k}}{\partial  n_{\beta,k'}}\right)^2 \!\!\frac{C_\beta(k')}{|v_{\alpha,k} -v_{\beta,k'}|}.
\end{align}
We derive explicit expressions for $C_\beta(k')$ below. We note that the geometric picture in principle allows us to compute higher-order corrections to propagation beyond diffusion (if we include the diffusive broadening in our estimate of the region over which fluctuations are computed), but we will not pursue these corrections here. 

Eq.~\eqref{front2} captures the diffusion of any type of quasiparticle.  To characterize the width of the ``front'' of a spreading operator, we simply compute the broadening of the quasiparticle with the largest velocity.

\emph{FFA model}.---We now explicitly check this result in the case of the Floquet-Fredrickson-Andersen model, an adaptation of Bobenko's Rule 54 cellular automaton~\cite{bobenko}. 
The diffusive broadening of operator fronts in this model was numerically demonstrated in Ref.~\cite{sg_ffa}. The dynamics of the model are given by the sequence of unitary gates
\begin{align}\label{unitary}
U = W(\text{odd} \rightarrow \text{even}) W(\text{even} \rightarrow \text{odd}),
\end{align}
where $W(\text{even} \rightarrow \text{odd})$ applies the following rule to each odd spin $n$: apply the Pauli operator $\sigma^x_n$ unless the neighboring even sites, $n-1$ and $n+1$, are both in the $\left| \downarrow \right.\rangle$ state; likewise for $W(\text{odd} \rightarrow \text{even})$ with even and odd sites interchanged.   These rules are implemented using standard quantum gates~\cite{sg_ffa}. The unit cell consists of two sites; in what follows we measure space in terms of unit cells. (The dynamics is symmetric under simultaneous spatial translation by a single lattice site and time translation by a half-step, but not under either operation separately.)

The dynamics of the FFA model can be described in terms of left- and right-moving quasiparticles. Each quasiparticle of either type has the same velocity, i.e., the dispersion relation is purely linear. This strict dispersionlessness is a distinctive feature of Floquet models, and cannot exist in a local lattice Hamiltonian. When two quasiparticles collide, each is delayed by one time-step; thus the model resembles a gas of hard rods with length $-1$. Microscopically, a ``free'' right-moving quasiparticle consists of two up spins, occupying an odd site and the even site to its right; a left-mover is similar, but occupies an odd site and the even site to its \emph{left}. A configuration of the form $\downarrow \uparrow \downarrow$ contains a left- and a right-mover on top of each other; such composites form during collisions. We compute coarse-grained densities of right/left movers $\rho_{R/L}$ by simply counting these configurations in the microscopic state.

In generic integrable systems, the density of quasiparticles with each rapidity is separately conserved, so the conservation laws of the model can all be understood in terms of the rapidity distribution of quasiparticles. In the FFA model, by contrast, there are only two velocities, so specifying the densities of left- and right-movers is  not enough to fix the conservation laws. The remaining conservation laws are the asymptotic \emph{spacings} between adjacent left- and right-movers. The broadening of the front couples to the velocities and not the spacings, so in what follows we ignore the spacings.

\emph{Velocity renormalization and diffusion in the FFA model}.---
In the FFA model, right- and left-movers move ballistically with the respective velocities $v_R$ and $v_L$, which depend on the densities as  
\begin{align}\label{velocity}
v_{R/L} = \pm 1 \mp \frac{2 \rho_{L/R}}{1+\rho_R + \rho_L}.
\end{align}
These formulas coincide with the prediction for a hard rod gas with effective length $a=-1$, and can also be derived in an elementary way~\cite{SupMat} as the left movers slow down the right movers and vice versa. 

\begin{figure}[tb]
\begin{center}
\begin{minipage}{0.23\textwidth}
\includegraphics[width = \linewidth]{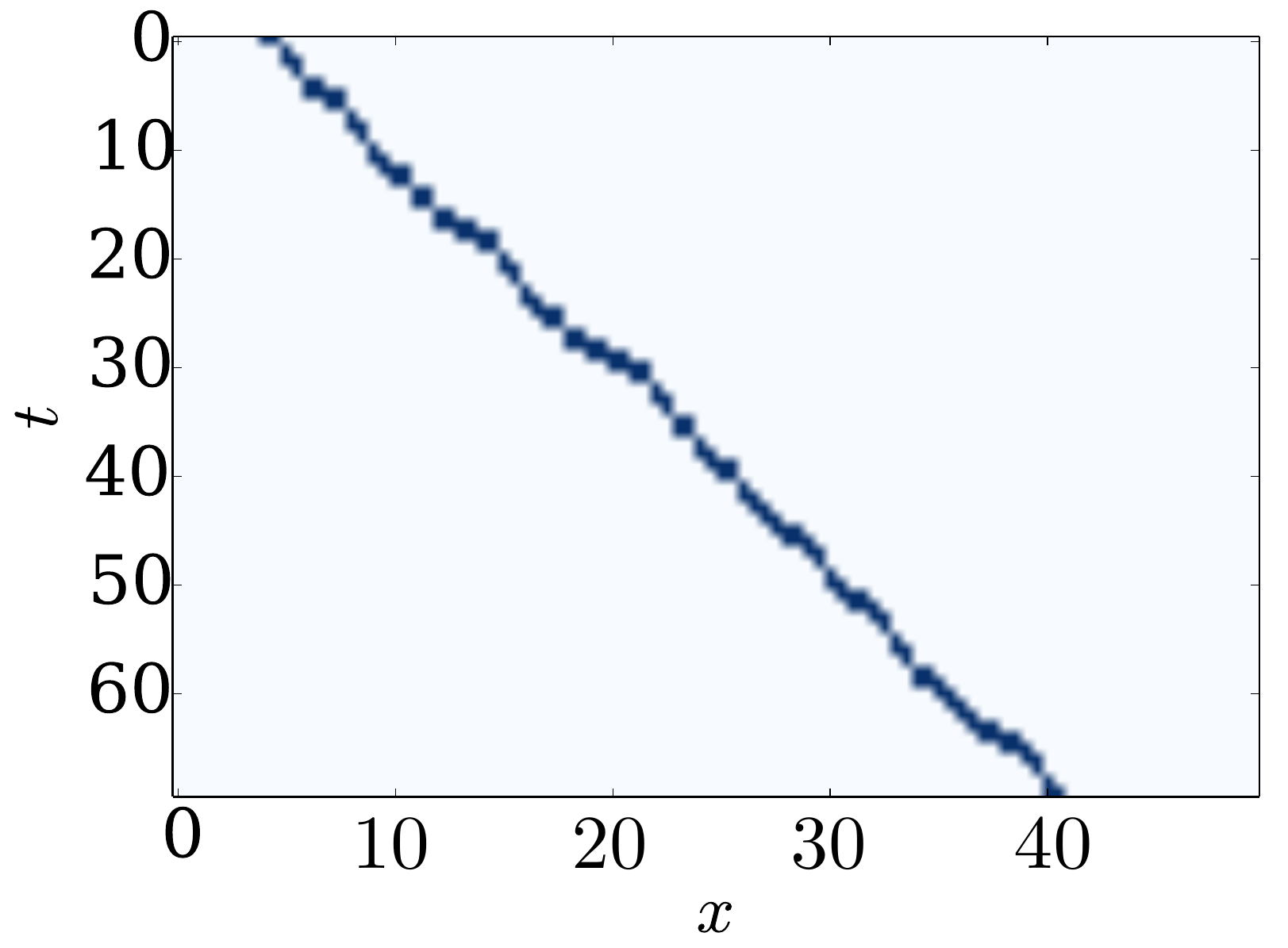}
\end{minipage}
\begin{minipage}{0.23\textwidth}
\includegraphics[width = \linewidth]{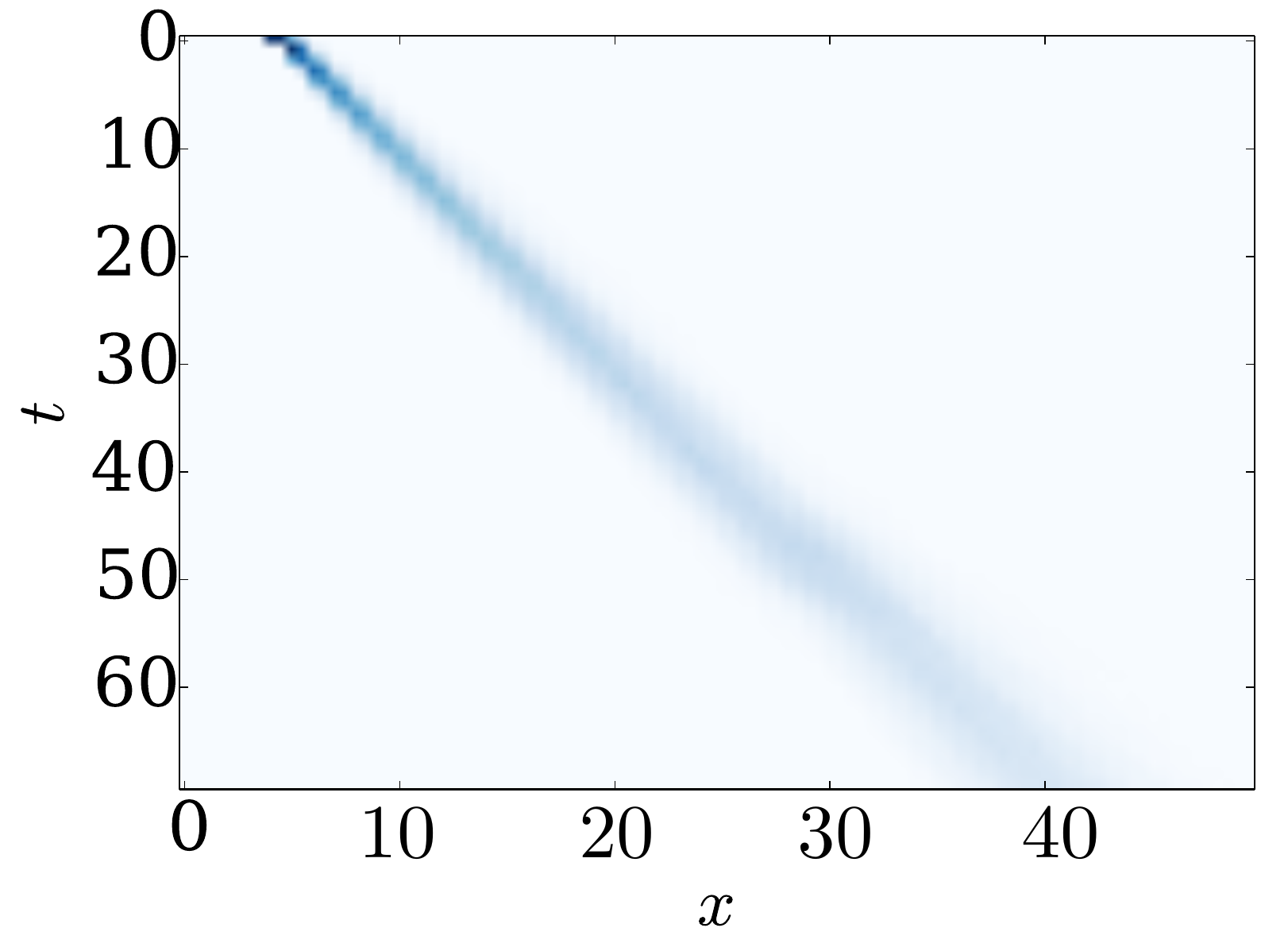}
\end{minipage}
\includegraphics[width = 0.45\textwidth]{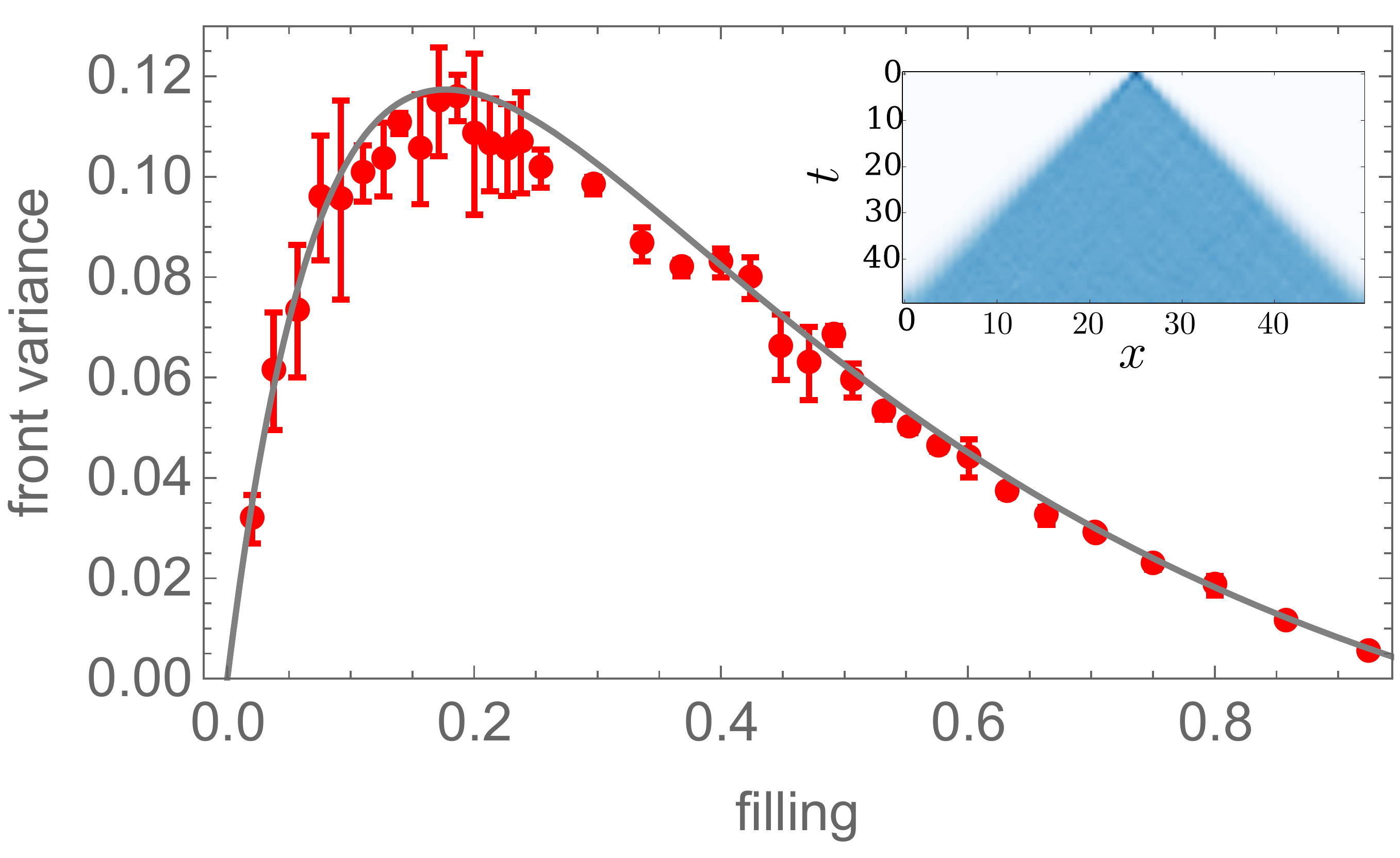}
\caption{Upper panel: biased random walk of a single right-moving quasiparticle in the FFA model for a specific initial product state (left) and diffusive broadening when averaged over 1000 product states (right). Lower panel: variance of front position vs. filling for a system of size $L=400$; numerical results (averaged over 1000 random product states) are in good agreement with the analytic formula~\eqref{ffaformula}. We emphasize that there are \emph{no} free parameters. Inset: OTOCs in the FFA model generically fill in behind the front~\cite{sg_ffa}.} 
\label{ffaresults}
\end{center}
\end{figure}

In order to include diffusive corrections, we incorporate equilibrium fluctuations of the quasiparticle densities, which lead to velocity fluctuations through Eq.~\eqref{velocity}. The density fluctuations are not diagonal in the left/right-mover basis, i.e., $\langle \delta \rho_L \delta \rho_R \rangle \neq 0$ with $\delta \rho_{R/L} = \rho_{R/L}  - \langle \rho_{R/L} \rangle  $. However, we can define the generalized ``Fermi factor'' $ n_{R/L} = 3 \rho_{R/L} / (1 + \rho_R + \rho_L)$; note that $\langle \delta  n_L \delta  n_R \rangle = 0$. The GHD equations describe the advection of these Fermi factors~\cite{SupMat}. From eq.~\eqref{velocity},  the velocity of a right-mover is given by $v_R = 1 - 2  n_L /3$; we have confirmed this expression numerically. (For simplicity we express our results for right-movers, but exactly analogous expressions can be written for left-movers.) To compute the velocity fluctuations, we need to compute the fluctuations $\langle \delta n_L \delta n_L \rangle$. We do this to leading order, by expanding $\delta  n_L $ in terms of $\delta \rho_R$ and $\delta \rho_L$, and computing the fluctuations of the densities. To do this we write a partition function $Z(\mu_L, \mu_R) = \sum\nolimits_{\{ \sigma \}} \exp(-\mu_L N_L - \mu_R N_R)$, and compute $Z$ using a $4\times 4$ transfer matrix, from which density fluctuations can be evaluated by taking derivatives~\cite{SupMat}. We specialize to $\rho_L = \rho_R = \rho$; in this case, the density fluctuations fit the analytic form $\langle (\delta  n_L)^2 \rangle = \frac{9 \rho (1 - \rho)}{(1 + 2 \rho)^4 \ell}$ for a system of size $\ell$. According to our kinetic argument, we compute the fluctuations over a distance $\ell = t \left|v_R - v_L \right| = 2 t v_R$.
Plugging these results into Eq.~\eqref{front}, we arrive at the following analytic expression for the variance of the quasiparticle position
\begin{align}\label{ffaformula}
\delta x^2(t) =  t \frac{2 \rho(1 - \rho)}{(1 + 2 \rho)^3},
\end{align}
for a tagged quasiparticle propagating through an equilibrium state with density (filling) $\rho$. This prediction is in good agreement with numerical simulation of the dynamics (Fig.~\ref{ffaresults}). The error bars indicate least-squares error in the fitting of the variance of the front. The small deviations from the theoretical prediction can be attributed to the early-time, sub-leading corrections that are relatively large in this regime -- the front is fitted after a few hundred time steps. The simplicity of our model allows us to directly measure the dynamics of a tagged quasiparticle, as follows. We evaluate the OTOC $\langle [P_\downarrow(i-1,B) \sigma^+(i,A) \sigma^+(i,B) \sigma^-(i+1,A) \sigma^-(i+1,B) P_\downarrow(i+2,A), \sigma^z_j(t)]^2\rangle$ where $P_{\uparrow/\downarrow}$ are projectors onto up/down spins, and $A$ and $B$ label the two spins in the $i$th unit cell. This corresponds to translating a single right-mover without creating or destroying any quasiparticles. In hard-rod models generally, translating a quasiparticle does not cause a butterfly effect; instead, the OTOC simply gives the time trace of the tagged quasiparticle (Fig.~\ref{ffaresults}). We emphasize that the existence of operators that can tag and translate single quasiparticles is a special feature of the FFA model; OTOCs of other generic local operators in the FFA model fill in behind the front and look similar to the chaotic case~\cite{sg_ffa} (Fig.~\ref{ffaresults}). 

In general interacting integrable models, a local operator that exclusively translates a single quasiparticle is unlikely to exist. Acting with a local operator generically at least changes the pseudomomenta of some quasiparticle(s) and thus the phase shifts of all the others~\cite{1742-5468-2016-6-063101,PhysRevB.95.115128}. Therefore, OTOCs of local operators should generically fill in behind the front.  Numerical results for the XXZ chain are shown in~\cite{SupMat}. Our numerics do not settle whether the saturation value behind the front in the XXZ model is distinct from the chaotic case, but the distinction, if present, is empirically weak for the operators we have considered.

\emph{Generic integrable systems}.--- Our picture can be straightforwardly generalized to other integrable systems with a quasiparticle description, like the XXZ spin chain. 
Such systems can be described in terms of stable quasiparticles, even at infinite temperature. 
Equilibrium states associated with a generalized Gibbs ensemble (GGE)~\cite{Rigol:2008kq,2016arXiv160403990V} correspond to a distribution of occupied quasiparticle states $\rho_{\alpha,k}$ with $\alpha$ the particle type and $k$ a pseudo-momentum~\cite{1742-5468-2016-6-063101,PhysRevB.95.115128}. After a quench, we expect the system to locally approach a GGE on a short timescale; our results apply for later times. Densities $\rho_k$ are related to the density of states $\rho^{\rm tot}_{\alpha,k}$ through the so-called Bethe equation  $\rho^{\rm tot}_{\alpha,k} +  \sum_\beta \int dk' \mathcal{K}_{\alpha \beta}(k,k')  \rho_{\beta,k'}  =  \partial_k p^0_{\alpha,k} /(2\pi)$, where the kernel $\mathcal{K}_{\alpha \beta}(k,k')$ encodes the two-body phase shifts of the model~\cite{Takahashi}, and $p^0_{\alpha,k}$ is the momentum. Ballistic transport in these models can be captured in terms of GHD, where the quasiparticle densities are assumed to be defined locally $\rho_{\alpha,k} (x,t)$. 
The effective quasiparticle velocity $v_{\alpha,k} [\rho]$ depends on the densities of all the other quasiparticles through some effective ``dressing'' operation by the interaction kernel $\mathcal{K}_{\alpha \beta}(k,k')$. 

Diffusive broadening is captured by reintroducing fluctuations in the GHD picture. In a GGE state, the fluctuations of the generalized Fermi factor $ n_{\alpha,k} = \rho_{\alpha,k}/\rho^{\rm tot}_{\alpha,k}$ in an interval of length $\ell$ are diagonal in $k$, and are given by~\cite{PhysRevB.54.10845} 
$\langle \delta  n_{\alpha,k} \delta  n_{\beta,k'} \rangle = \delta_{\alpha,\beta} \delta(k- k')   n_{\beta,k'}  (1- n_{\beta,k'} ) / (\rho^{\rm tot}_{\beta,k'} \ell).$ 
(Note that $n_k$ is a dimensionless quantity.) In the basis of generalized Fermi factors, thermal fluctuations are essentially free-fermion-like. 
From the explicit form of the quasiparticle velocity in terms of generalized Fermi factors, we find the functional derivative~\cite{SupMat,1751-8121-50-43-435203} 
$ \rho^{\rm tot}_{\alpha,k} \delta v_{\alpha,k} / \delta  n_{\beta,k'} = ( v_{\alpha,k} -  v_{\beta,k'}) \rho^{\rm tot}_{\beta,k'} \mathcal{K}^{\rm dr}_{\alpha \beta}(k,k')$,  
where $\mathcal{K}^{\rm dr}_{\alpha \beta}(k,k')$ is a ``dressed'' version of the scattering kernel, which satisfies the integral equation $ \mathcal{K}^{\rm dr}_{\alpha \beta}(k,k') = \mathcal{K}_{\alpha \beta}(k,k') - \sum_{\gamma} \int dk'' \mathcal{K}_{\alpha \gamma}(k,k'') \mathcal{K}^{\rm dr}_{\gamma \beta}(k'',k')  n_{\gamma,k''}$.  Plugging this expression into~\eqref{front}, with the explicit form of the Fermi factor fluctuations computed over a distance $\ell = \left|v_{\alpha,k} -v_{\beta,k'} \right| t$, we find 
\begin{align}\label{broadeningGeneral}
\delta x^2_{\alpha,k} (t) &= t  \frac{1}{(\rho^{\rm tot}_{\alpha,k} )^2} \sum_\beta \int dk'  \left| v_{\alpha,k} -  v_{\beta,k'} \right|  \notag \\
& \times [ \mathcal{K}^{\rm dr}_{\alpha \beta}(k,k') ]^2 \rho_{\beta,k'} (1- n_{\beta,k'}) .
\end{align}
This formula gives an explicit expression for the diffusive broadening of a quasi-particle $\alpha$ with pseudo-momentum $k$ propagating through an homogenous equilibrium state due to thermal fluctuations. It can be evaluated explicitly for any integrable model, and for the fastest quasiparticle, coincides with the diffusive broadening of the operator spreading front.  

\emph{Transport}.---In integrable systems, unlike chaotic ones, the subleading diffusive quasiparticle spreading affects not just OTOCs but also time-ordered correlators and transport properties. To see this, consider a GGE state characterized by a generalized Fermi factor distribution $ n^{\rm eq}_{\alpha,k}$. A small perturbation $\hat  n_{\alpha,k} (x,t)$ over this GGE state propagates with mean velocity $v_{\alpha,k} [ n^{\rm eq}]$, but with diffusive broadening $\delta x^2_{\alpha,k} (t) = 2 D_k t$ given by eq.~\eqref{broadeningGeneral}. The corresponding linear-response hydrodynamic equation reads
\begin{align}\label{NavierStokes}
\partial_t \hat  n_{\alpha,k} + v_{\alpha,k} [ n^{\rm eq}]  \partial_x \hat  n_{\alpha,k} =  D_k [ n^{\rm eq}]  \partial^2_x \hat  n_{\alpha,k}+\dots
\end{align}
where the dots include higher-derivative corrections, but also $ \partial^2_x \hat  n_{\beta,k'} $ terms with $(\beta,k') \neq (\alpha,k)$. In fact, although our derivations seem quite distinct, our expression for the diagonal diffusion constant $D_k$ coincides with the very recent ``Navier-Stokes'' corrections computed in Ref.~\cite{denardis2018} from a form-factor (matrix-element) expansion of the Kubo formula; it will be interesting to extend our argument to reproduce fully the transport equation of~\cite{denardis2018}. 

We note that although fronts for OTOCs and time-ordered correlation functions both broaden diffusively, there are important differences: 
First, operator spreading is dominated by the \emph{fastest} quasiparticle, whereas transport generally involves all quasiparticles. Second, OTOCs saturate to a non-zero value behind the front, while time-ordered correlators decay. Third, different conserved quantities couple differently to quasiparticles, leading to distinct transport properties (e.g., spin transport in the XXZ model is sub-ballistic for $J_z \geq J_x=J_y$~\cite{PhysRevLett.106.217206,PhysRevLett.119.020602}) so time-ordered correlators may not detect the ballistic operator front in all cases. Numerical results for two-point correlators, OTOCs, and diffusive front broadening for the XXZ model are shown in~\cite{SupMat}. 

\emph{Discussion}.---This paper used a simple, physically motivated picture from kinetic theory to derive diffusive corrections to ballistic quasiparticle spreading in interacting integrable systems, and its implications for operator spreading and transport. We showed that OTOCs in interacting integrable models have diffusive front broadening just as in non-integrable systems, although the mechanisms are quite different. Nevertheless, coarse grained measures of front dynamics are not able to discriminate between these mechanisms. 
Whether our hydrodynamic approach can be generalized to construct a fluctuating hydrodynamics of integrable systems, or extended to situations like the isotropic Heisenberg chain for which the diffusion is anomalous~\cite{Ljubotina:2017aa,2018arXiv180603288I}, are left for future work.

In addition to the diffusive effect discussed here, interacting integrable systems have a subleading $t^{1/3}$ front-broadening that they share with free-fermion models~\cite{Platini2005}. One might wonder if there are any natural circumstances in which the diffusive broadening we predict might be absent, causing the $t^{1/3}$ effect to dominate. Eq.~\eqref{broadeningGeneral} suggests that this essentially never happens in an interacting system, as all terms are non-negative and therefore the integrand would have to vanish identically (which is plausible in zero-entropy states, such as the ground state, but not otherwise).

\emph{Acknowledgments}.---The authors thank Denis Bernard, Vir Bulchandani, Pasquale Calabrese, Anushya Chandran, Andrea De Luca, Jacopo De Nardis, Benjamin Doyon, Adam Nahum, Vadim Oganesyan, Tomaz Prosen, and Brian Swingle for helpful discussions and comments on the manuscript. This work was supported by NSF Grant No. DMR-1653271 (S.G.), DOE grant No. DE-SC0016244 (D.A.H.), and US Department of Energy, Office of Science, Basic Energy Sciences, under Award No. DE-SC0019168 (R.V.). V.K. is supported by the Harvard Society of Fellows and the William F. Milton Fund. S.G., V.K. and R.V. are grateful to the KITP, which is supported by the National Science Foundation under Grant No. NSF PHY-1748958, and the program `The Dynamics of Quantum Information', where part of this work was completed. S.G. performed this work in part at the Aspen Center for Physics, which is supported by NSF Grant No. PHY-1607611.

\bibliography{References}

%

%

\bigskip

\onecolumngrid
\newpage



\section*{Supplementary material}

# Supplemental Material for "Hydrodynamics of operator spreading and quasiparticle diffusion in interacting integrable systems"

Sarang Gopalakrishnan,[1] David A. Huse,[2] Vedika Khemani,[3] and Romain Vasseur[4]

[1]*Department of Physics and Astronomy, CUNY College of Staten Island,*
*Staten Island, NY 10314; Physics Program and Initiative for the Theoretical Sciences,*
*The Graduate Center, CUNY, New York, NY 10016, USA*
[2]*Physics Department, Princeton University, Princeton, New Jersey 08544, USA*
[3]*Department of Physics, Harvard University, Cambridge, MA 02138, USA*
[4]*Department of Physics, University of Massachusetts, Amherst, MA 01003, USA*
(Dated: November 19, 2018)

## I. CONTINUITY EQUATIONS IN THE FFA MODEL

The densities or right and left moving quasiparticles in the FFA models can be written as

$$\begin{aligned}
\rho_R(i) &= P_\uparrow(i,A)P_\uparrow(i,B) + P_\downarrow(i-1,B)P_\uparrow(i,A)P_\downarrow(i,B) + P_\downarrow(i,A)P_\uparrow(i,B)P_\downarrow(i+1,A), \\
\rho_L(i) &= P_\uparrow(i-1,B)P_\uparrow(i,A) + P_\downarrow(i-1,B)P_\uparrow(i,A)P_\downarrow(i,B) + P_\downarrow(i,A)P_\uparrow(i,B)P_\downarrow(i+1,A),
\end{aligned} \quad (1)$$

where $N_R = \sum_i \rho_R(i)$ and $N_L = \sum_i \rho_L(i)$ are conserved by the time evolution. It is convenient to define

$$\begin{aligned}
\rho_+(i) &= P_\uparrow(i,A)P_\uparrow(i,B) + P_\uparrow(i-1,B)P_\uparrow(i,A) + 2P_\downarrow(i-1,B)P_\uparrow(i,A)P_\downarrow(i,B) \\
&\quad + P_\downarrow(i,A)P_\uparrow(i,B)P_\downarrow(i+1,A) + P_\downarrow(i-1,A)P_\uparrow(i-1,B)P_\downarrow(i,A), \\
\rho_-(i) &= P_\uparrow(i,A)P_\uparrow(i,B) - P_\uparrow(i,B)P_\uparrow(i+1,A).
\end{aligned} \quad (2)$$

The sum of these densities is also conserved, and upon coarse graining, we have $\rho_\pm(x) = \rho_R(x) \pm \rho_L(x)$. Given the simple dynamics of the FFA model, one can readily compute the time evolution of these operators over a single Floquet cycle. For example, $U^\dagger(P_\uparrow(i,A)P_\uparrow(i,B))U = P_\downarrow(i-1,B)P_\uparrow(i,A)P_\downarrow(i,B) + P_\uparrow(i-1,B)P_\uparrow(i,A)P_\downarrow(i,B)$, and $U^\dagger(P_\downarrow(i-1,B)P_\uparrow(i,A)P_\downarrow(i,B))U = P_\uparrow(i-1,B)P_\downarrow(i,A)P_\uparrow(i,B)$. Using these relations and rewriting $\rho_\pm(i,t+1) - \rho_\pm(i,t)$ as telescopic sums, we find, after a lengthy but straightforward calculation

$$\begin{aligned}
\rho_+(i,t+1) - \rho_+(i,t) + j_+(i+1,t) - j_+(i,t) &= 0, \\
\rho_-(i,t+1) - \rho_-(i,t) + j_-(i+1,t) - j_-(i,t) &= 0.
\end{aligned} \quad (3)$$

The lattice currents are given by

$$\begin{aligned}
j_+(i) &= \rho_-(i-1), \\
j_-(i) &= P_\downarrow(i-1,B)P_\uparrow(i,A)P_\downarrow(i,B) + P_\uparrow(i-1,B)P_\downarrow(i,A) + P_\downarrow(i-1,A)P_\downarrow(i,A)P_\uparrow(i,B).
\end{aligned} \quad (4)$$

The first equation indicates that the total current $J_+ = \sum_i j_+(i)$ is a conserved quantity, implying ballistic transport for $\rho_+$.

## II. THERMODYNAMICS AND HYDRODYNAMICS OF THE FFA MODEL

Equilibrium states in the FFA model can be characterized by the classical partition function

$$Z = \sum_{\{\sigma\}} e^{-\mu_R N_R - \mu_L N_L}, \quad (5)$$

which can be computed using the $4 \times 4$ transfer matrix

$$T = \begin{bmatrix} 1 & 1 & e^{-\mu_R-\mu_L} & e^{-\mu_R/2} \\ e^{-\mu_R-\mu_L} & e^{-\mu_R-\mu_L} & e^{-\mu_L} & e^{-\mu_R/2-\mu_L} \\ 1 & 1 & e^{-\mu_R-\mu_L} & e^{-\mu_R/2} \\ e^{-\mu_R/2} & e^{-\mu_R/2} & e^{-\mu_R/2-\mu_L} & e^{-\mu_R-\mu_L} \end{bmatrix}, \quad (6)$$

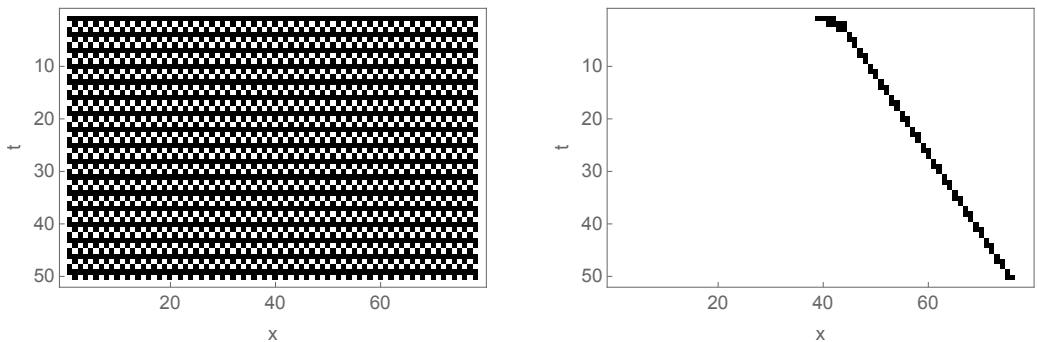

FIG. 1: Nature of fluctuations and transport in the FFA model near full filling. The dynamics of the model is oscillatory with period 3 (left); thus, quasiparticles move with a velocity that also oscillates with period 3 (right). After coarse-graining in time, though, the quasiparticles can be seen to move exactly ballistically with no front-broadening.

written in the basis $\{\downarrow\downarrow, \uparrow\downarrow, \downarrow\uparrow, \uparrow\uparrow\}$. Density averages and fluctuations in the thermodynamic limit can be obtained from various derivatives of the largest eigenvalue of $T$. We find that the density fluctuations are non-diagonal, $\langle\delta\rho_R\delta\rho_L\rangle \neq 0$ in the right and left moving basis. However, if define the generalized "Fermi factor" $n_{R/L} = 3\rho_{R/L}/(1+\rho_R+\rho_L)$, we find that

$$\langle(\delta n_L)^2\rangle = \langle(\delta n_R)^2\rangle = \frac{9\rho(1-\rho)}{(1+2\rho)^4 \ell},$$
$$\langle\delta n_R \delta n_L\rangle = 0, \tag{7}$$

for a system of size $\ell$ in an equilibrium state with filling $\rho = \rho_R = \rho_L$.

Using this transfer matrix, we can also compute the equilibrium averages of the currents $j_\pm = \rho_R v_R \pm \rho_L v_L$. We find

$$v_{R/L} = \pm 1 \mp \frac{2\rho_{L/R}}{1+\rho_R+\rho_L}. \tag{8}$$

This formula coincides with the velocities of a hard rod gas with only two "bare" velocities $v_R^0 = 1$ and $v_L^0 = -1$, and effective length $a = -1$.[1,2] One can also understand this formula in an elementary way as follows: if a right mover starts at $(0,0)$ and goes to $(x,t)$, it encounters all the left movers that started at points closer than $x - v_L t$. The time $t$ is determined by accounting for the fact that the "bare" velocity is 1, and that each collision with a left mover adds an extra time step. This means that $t = x + \rho_L(x - v_L t)$, which yields $v_R = x/t = (1 + \rho_L v_L)/(1 + \rho_L)$. Repeating the argument for left movers, we find $v_L = (-1 + \rho_R v_R)/(1 + \rho_R)$. Solving these two equations yields (8). Note that this formula is consistent with

$$j_+ = \rho_R - \rho_L, \tag{9}$$

as it should since this relation holds exactly microscopically.

Since the microscopic averages of the currents coincide with this hydrodynamic expectation, one might be tempted to use the variance of the microscopic current instead of the expression we obtained from the density fluctuations and the coarse-grained velocity formulas (8). These two expressions do *not* agree; the discrepancy between them can be seen, and its mechanism most clearly understood, near the fully filled limit. At full filling, the dynamics consists of persistent period-3 oscillations (Fig. 1); the current oscillates in a 100100100... sequence. These current fluctuations, though strong, are not *random*, and they cause deterministic rather than random changes of the front velocity. After time averaging, these fluctuations cancel out exactly except for boundary terms. Thus they are not responsible for diffusive broadening. Instead, the fluctuations that cause diffusive broadening are the slow, random fluctuations that survive under coarse-graining.

The Euler hydrodynamics in the quasiparticle language (corresponding to the kinetic theory of these quasiparticles), leads to the following continuity equations

$$\partial_t \rho_{R/L} + \partial_x (v_{R/L}[\rho_R, \rho_L]\rho_{R/L}) = 0. \tag{10}$$

Using the definition of the "Fermi factors" above, we find the advection equations

$$\partial_t n_{R/L} + v_{R/L}[n_R, n_L]\partial_x n_{R/L} = 0. \tag{11}$$

This is a general property of Fermi factors in generic integrable systems.[3,4]

## III. QUASIPARTICLE DIFFUSION IN GENERIC INTEGRABLE MODELS

In this section, we derive explicitly the form of the quasiparticle diffusion using the kinetic theory argument given in the main text. The hydrodynamic equations for quantum integrable systems can be thought of either as Euler equations for all the (local and quasi-local) conserved quantities, or equivalently, as a kinetic equation for the quasi-particles. Let us define densities of *particles*, *holes* and *states* via

$$\ell \rho_k dk = \{\# \text{ occupied pseudo-momenta in } [k, k+dk]\}$$
$$\ell \rho_k^h dk = \{\# \text{ unoccupied pseudo-momenta in } [k, k+dk]\}$$
$$\ell \rho_k^{\text{tot}} dk = \{\# \text{ allowed pseudo-momenta in } [k, k+dk]\}. \tag{12}$$

respectively, for a system of size $\ell$. For simplicity, we are using a shorthand notation for $k$ which labels both the types of quasiparticles and their pseudo-momenta. Let $\rho_k^{\text{tot}} = \rho_k + \rho_k^h$ be the total density of states, and it is also useful to define the generalized Fermi factor, given by $n_k = \frac{\rho_k}{\rho_k^{\text{tot}}}$, which is by definition a number between 0 and 1. These quantities are related by the *Bethe equations*

$$\rho_k^{\text{tot}} + \int_{-\infty}^{\infty} \mathcal{K}(k, k') \rho_{k'} dk' = \frac{\partial_k p_k^0}{2\pi}, \tag{13}$$

for some scattering kernel $\mathcal{K}(k, k')$ whose precise form depends on the integrable model under consideration, with $p_k^0$ the bare momentum. A given generalized Gibbs ensemble (GGE) state – including thermal states in particular – can be shown to correspond to a given distribution of quasi-particles $\rho_k$. Note that we can also choose to work with the Fermi factor $n_k$, since the Bethe equation together with $\rho_k^{\text{tot}} = \rho_k + \rho_k^h$ fully fix the other distributions once $n_k$ is known.

Assuming local equilibrium, we can imagine that these quantities all depend on $x$ and $t$, and that the Bethe equation (13) is satisfied locally. The semi-classical kinetic equation for $n_k$ is then quite simple[3,4]

$$\partial_t n_k + v_k[n] \partial_x n_k = 0, \tag{14}$$

where $v_k[n]$ is the group velocity of the quasi-particles which can also be computed from Bethe ansatz. The effective group velocity[5] is then given by $v_k[n] = \frac{\epsilon'_k[n]}{p'_k[n]}$, where $\epsilon'_k$ and $p'_k$ are the dressed derivatives of the quasiparticle energy and momentum, given by

$$\epsilon'_k + \int dk' \mathcal{K}(k - k') n_{k'} \epsilon'_{k'} = \partial_k \epsilon_k^0, \tag{15}$$

$$p'_k + \int dk' \mathcal{K}(k - k') n_{k'} p'_{k'} = \partial_k p_k^0. \tag{16}$$

with $\epsilon_k^0$ and $p_k^0$ the bare energy and momentum. Note that these equations are time-reversal invariant and ignore the density fluctuations responsible for diffusion that we now describe.

For a system of size $\ell$, the fluctuations of the quasiparticle densities in a thermal state at temperature $T$ (corresponding to an equilibrium Fermi factor $n_k$ given by the thermodynamic Bethe ansatz)[6] read[7]

$$\langle \delta n_k \delta n_{k'} \rangle = \frac{1}{\rho_k^{\text{tot}} \ell} n_k (1 - n_k) \delta(k - k'). \tag{17}$$

As explained in the main text, these fluctuations should be computed over a distance $\ell = |v_k - v_{k'}|t$. This leads to a broadening of the quasiparticle trajectories

$$\delta x_k^2(t) = t \int dk' \left(\frac{\delta v_k}{\delta n_{k'}}\right)^2 \frac{n_{k'}(1 - n_{k'})}{\rho_{k'}^{\text{tot}} |v_k - v_{k'}|}. \tag{18}$$

The last step of the calculation is to compute the functional derivative $\frac{\delta v_k}{\delta n_{k'}}$ using eqs. (15) and (16). We see that we have $\frac{\delta \epsilon'_k}{\delta n_{k'}} = v_{k'} \frac{\delta p'_k}{\delta n_{k'}}$, so that

$$\frac{\delta v_k}{\delta n_{k'}} = \frac{v_{k'} - v_k}{p'_k} \frac{\delta p'_k}{\delta n_{k'}} = \frac{v_{k'} - v_k}{\rho_k^{\text{tot}}} \frac{\delta \rho_k^{\text{tot}}}{\delta n_{k'}}, \tag{19}$$

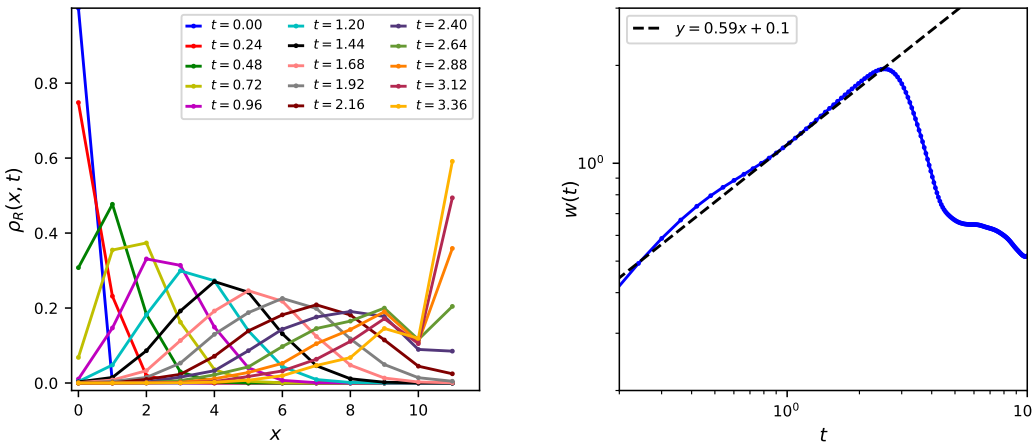

FIG. 2: (left) Right weight $\rho_R(x,t)$ for the spreading operator $\sigma_0^x(t)$ in an XXZ chain of size $L = 12$ with $\Delta = 0.5$ showing ballistic front propagation with front broadening. Width of the spreading front $w(t)$ shows a scaling consistent with $w(t) \sim t^{1/2}$ (dashed line: fit with $x = \log t$ and $y = \log w$). Note that we have very limited dynamical range in time before the front reaches the end of the chain, but the numerical exponent is already quite far from the free fermion result $w(t) \sim t^{1/3}$.

since $p'_k = 2\pi\rho_k^{\text{tot}}$ (see also Refs.8,9 for related formulas). Let $-\frac{1}{p'_{k'}} \frac{\delta p'_k}{\delta n_{k'}} \equiv \mathcal{K}^{\text{dr}}(k,k')$. Using (16), this "dressed kernel" satisfies the integral equation

$$\mathcal{K}^{\text{dr}}(k,k') = \mathcal{K}(k,k') - \int dk'' \mathcal{K}(k,k'') \mathcal{K}^{\text{dr}}(k'',k') n_{k''}. \tag{20}$$

In terms of the dressed kernel, we have $\left(\frac{\delta v_k}{\delta n_{k'}}\right)^2 = (v_k - v_{k'})^2 \left[\mathcal{K}^{\text{dr}}(k,k')\right]^2 \left(\frac{\rho_{k'}^{\text{tot}}}{\rho_k^{\text{tot}}}\right)^2$. Plugging this expression in (18), we find

$$\delta x_k^2(t) = t \frac{1}{(\rho_k^{\text{tot}})^2} \int dk' \, |v_k - v_{k'}| \left[\mathcal{K}^{\text{dr}}(k,k')\right]^2 \rho_{k'}(1 - n_{k'}), \tag{21}$$

as claimed in the main text. This formula can be evaluated in any equilibrium (GGE) state.

## IV. NUMERICS ON THE XXZ SPIN CHAIN

We now present some numerics on the XXZ chain using exact diagonalization. Despite the small sizes and times accessible to this study, we see many signatures of the qualitative features discussed in the main text, including the diffusive front broadening of operators.

The Hamiltonian is

$$H = \sum_i \sigma_i^x \sigma_{i+1}^x + \sigma_i^y \sigma_{i+1}^y + \Delta \sigma_i^z \sigma_{i+1}^z \tag{22}$$

where $\sigma_i^{x/y/z}$ are Pauli spin 1/2 operators on site $i$ and we pick $\Delta = 0.5$ for specificity. For a spin-1/2 chain of length $L$, a complete orthonormal basis for all operators is given by the $4^L$ "Pauli strings" $\mathcal{S}$, which are products of Pauli matrices on distinct sites. We can then express our spreading operator in this basis of Pauli strings:

$$O_0(t) = \sum_{\mathcal{S}} a_{\mathcal{S}}(t) \mathcal{S} . \tag{23}$$

We measure the right front of the spreading operator $O(t)$ using the "right-weight" $\rho_R(x,t)$, defined as the total weight in $O(t)$ of basis strings that end at site $x$ — which means that they act as the identity on all sites to the right of site $x$, but act as a non-identity on site $x$:

$$\rho_R(x,t) = \sum_{\substack{\text{strings } \mathcal{S} \text{ with} \\ \text{rightmost non-} \\ \text{identity on site } x}} |a_{\mathcal{S}}|^2, \qquad \sum_i \rho_R(x,t) = 1. \tag{24}$$

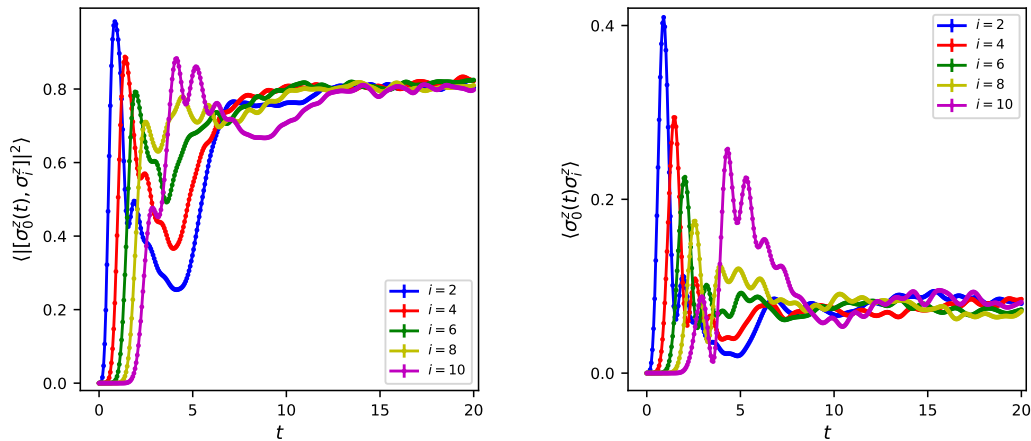

FIG. 3: (left) Two-point time-ordered correlators and OTOCs in the XXZ model at $\Delta = 0.5$ showing that both these quantities detect the operator front(s), although they saturate to different values at late times. The multiple fronts visible in the data is consistent with the presence of multiple species of quasiparticles.

The conservation law on $\rho_R(x,t)$ follows from unitarity and supports the identification of this quantity as an "emergent" density undergoing biased diffusion.[10,11] The left front can be defined analogously. Fig 2 (right) shows $\rho_R(x,t)$ for an initial operator $O_0 = \sigma_0^x$ at different times in a system of length $L = 12$. We clearly see front propagating ballistically to the right as the weight of $O(t)$ spreads to longer Pauli strings; we also clearly see the front getting broader with increasing time before it reaches the end of the chain. We can quantify the width of the front by looking at the second moment of $\rho_R(x,t)$ in time:

$$w(t) = \sqrt{\sum_x x^2 \rho_R(x,t) - \left(\sum_x x \rho_R(x,t)\right)^2}. \qquad (25)$$

Fig 2 (left) shows $w(t)$ plotted against time. While we have very limited dynamical range in time before the front reaches the end of the chain (and after an initial transient), the data seems consistent with a $w(t) \sim t^{1/2}$ scaling, and certainly quite far from the free fermion result of $w(t) \sim t^{1/3}$. The data for the spreading of $\sigma_0^z(t)$ looks very similar.

Next, we look at the behavior of two-point correlators and OTOCs: $\langle \sigma_0^z(t) \sigma_i^z(t) \rangle$ and $\langle |[\sigma_0^z(t), \sigma_i^z(t)]|^2 \rangle$ in Fig 3. We see that *both* the two-point correlator and the OTOC are able to detect the operator front for integrable systems, unlike the case of non-integrable systems (although two-point functions may not always work for this purpose, for example when additional symmetries force these to be zero or subballistic). Notice that the figures show the appearance of multiple fronts which is consistent with the presence of multiple species of quasiparticles in the XXZ model, each with their own "fastest" speed. Further, as discussed in the main text, while the two point function decays at late times, the OTOC generically saturates to a non-zero value as the front fills in. The saturation value of the OTOC is approximately 0.8 which is less than the generic value of 1 in chaotic systems. However, this difference is small and the saturation does show a weak increase with increasing system size (not shown). Whether this saturation value converges to a distinct universal value in large systems is presently unclear to us, but the difference is certainly weak. We note also that the non-monotonic behavior of the OTOC before saturation is generally not seen in chaotic models, so this may again be a weak signature of the lack of chaos. But the origin and universality of this behavior is again presently unsettled.



\end{document}